\documentclass[twocolumn,aps,pra,10pt,superscriptaddress,letterpaper,citeautoscript,showpacs,nobibnotes]{revtex4-1}

\usepackage{amsfonts,amsmath,amssymb}

\usepackage{bm}

\usepackage[dvipdfm]{graphicx}

\usepackage{color}

\makeatletter
\renewcommand{\@pacs@name}{Subject Areas: }
\makeatother

\def\usedvipdfmx{0} 
\def\inclrefttl{1} 

\graphicspath{{./FigureMain/}{./FigureSuppInfo/}}

\if \usedvipdfmx 1
\else
\fi

\if \inclrefttl 1
  \def\citetitle#1{{\it #1},}
\else
  \newcommand{\citetitle}[1]{}
\fi

\newcommand{\nature}{ Nature (London) }
\newcommand{\natphoton}{Nat.\ Photon.\ }
\newcommand{\natphys}{Nat.\ Phys.\ }
\newcommand{\natmater}{Nat.\ Mater.\ }
\newcommand{\science}{Science }
\renewcommand{\prl}{Phys.\ Rev.\ Lett.\ }
\renewcommand{\pra}{Phys.\ Rev.\ A }
\renewcommand{\rmp}{Rev.\ Mod.\ Phys.\ }
\newcommand{\optexp}{Opt.\ Exp.\ }
\newcommand{\njp}{New J.\ Phys.\ }
\newcommand{\epjd}{Eur.\ Phys.\ J.\ D }

\newcommand{\epjap}{Eur.\ Phys.\ J.\ Appl.\ Phys.\ }
\newcommand{\rsi}{Rev.\ Sci.\ Instrum.\ }
\newcommand{\mst}{Meas.\ Sci.\ Technol.\ }
\newcommand{\wirescs}{WIREs Comp Stat }

\newcommand{\qph}[1]{arXiv:#1 [quant-ph] }

\newcommand{\etal}{{\it et al.,}}

\renewcommand{\Re}{\mathrm{Re}}
\renewcommand{\Im}{\mathrm{Im}}
\newcommand{\natnum}{\mathbb{N}}
\newcommand{\ket}[1]{ |{#1}\rangle }
\newcommand{\bra}[1]{ \langle{#1}| }

\DeclareMathOperator{\Tr}{Tr}

\newcommand{\UTokyo}{Department of Applied Physics, School of Engineering, \\
The University of Tokyo, 7-3-1 Hongo, Bunkyo-ku, Tokyo 113-8656, Japan}
\newcommand{\UMainz}{Institute of Physics, Staudingerweg 7,
Johannes Gutenberg-Universit\"{a}t Mainz, 55099 Mainz, Germany}

\begin{document}

\title{Creation, Storage, and On-Demand Release \\
of Optical Quantum States with a Negative Wigner Function}

\author{Jun-ichi Yoshikawa}
\email{yoshikawa@ap.t.u-tokyo.ac.jp}
\affiliation{\UTokyo}
\author{Kenzo Makino}
\affiliation{\UTokyo}
\author{Shintaro Kurata}
\affiliation{\UTokyo}
\author{Peter van Loock}
\affiliation{\UMainz}
\author{Akira Furusawa}
\email{akiraf@ap.t.u-tokyo.ac.jp}
\affiliation{\UTokyo}

\date{\today}

\begin{abstract}

Highly nonclassical quantum states of light, characterized by Wigner functions with negative values,
have been created so far only in a heralded fashion.
In this case, the desired output emerges rarely and randomly from a quantum-state generator.
An important example is the heralded production of high-purity single-photon states, typically based
on some nonlinear optical interaction.
In contrast, on-demand single-photon sources were also reported, exploiting the quantized level structure of matter systems.
These sources, however, lead to highly impure output states, composed mostly of vacuum.
While such impure states may still exhibit certain single-photon-like features such as anti-bunching, they are not enough nonclassical for advanced quantum information processing.
On the other hand, the intrinsic randomness of pure, heralded states can be circumvented by first storing and then releasing them on demand.
Here we propose such a controlled release, and we experimentally demonstrate it for heralded single photons.
We employ two optical cavities, where the photons are both created and stored inside one cavity, and finally released through a dynamical tuning of the other cavity.
We demonstrate storage times of up to $300$ ns, while keeping the single-photon purity around $50\%$ after storage.
This is the first demonstration of a negative Wigner function at the output of an on-demand photon source or a quantum memory.
In principle, our storage system is compatible with all kinds of nonclassical states, including those known to be essential for many advanced quantum information protocols.

\end{abstract}

\pacs{Quantum Information, Quantum Physics, Photonics}

\maketitle

\section{INTRODUCTION}

Representing flying quantum information, photons are ideally suited for communication between the stations of a quantum network \cite{Gisin.NatPhoton(2007),Kimble.Nature(2008)}.
Among various photonic quantum states, single-photon states are the most fundamental. In principle, single-photon-based qubits would allow for scalable quantum computation based on linear-optical circuits, auxiliary quantum states of single and many photons, and photon counters \cite{Knill.Nature(2001),Kok.RMP(2007)}.
However, nonclassical states beyond single photons are also expected to be useful for future advanced quantum information processing \cite{Mari.PRL(2012),Gottesman.PRA(2001),Marek.PRA(2011),Bimbard.NatPhoton(2010),Yukawa.OptExp(2013),Yukawa.arXiv(2013),Lee.Science(2011),Takeda.Nature(2013)}.

Single-photon sources are often divided into two types \cite{Eisaman.RSI(2011)}, where one corresponds to transitions in electronic energy levels accompanied by a photon emission, and the other is implemented by probabilistically generating photon pairs, and detecting one photon and thus heralding another one.
The first type of sources allow for an on-demand emission of photons, which has been demonstrated with various matter systems \cite{Eisaman.RSI(2011),Buller.MST(2010)}, such as trapped atoms, defects in diamonds, and semiconductor quantum dots.
However, these systems typically share a common disadvantage, namely a very low efficiency for collecting the emitted photons in a specific single spatial mode, and so it remains unclear whether a photon emission has actually happened until the photon is detected \cite{Buller.MST(2010)}.
So far, only the class of heralded sources offers high fidelity and flexibility, though at the expense of their intrinsic randomness.

The photon pairs of the heralded sources are typically produced via an optical nonlinearity, namely either parametric down conversion \cite{Castelletto.EPJAP(2008)} or four-wave mixing \cite{Fulconis.PRL(2007)}.
Since this only requires a perturbative coupling between light and matter, heralded photons can be emitted in a well-defined mode with good purity and optical coherence.
As a consequence, a negative Wigner function has been observed so far only with this type of source \cite{Lvovsky.PRA(2001),Neergaard-Nielsen.OptExp(2007)}.
Moreover, measuring parts of a nonlinear optical resource state also allows for the heralded preparation of quantum states beyond single photons or qubits \cite{Gottesman.PRA(2001),Marek.PRA(2011),Bimbard.NatPhoton(2010),Yukawa.OptExp(2013),Yukawa.arXiv(2013)}.
Finally, in contrast to the highly restrictive wavelengths of existing on-demand sources, basically determined by the corresponding matter energy levels, the heralded approach allows for a broader range of wavelengths, in particular, including the telecom wavelength \cite{Fasel.NJP(2004)}.
Owing to these advantages, the most advanced quantum information experiments have been demonstrated with such heralding schemes \cite{Bimbard.NatPhoton(2010),Yukawa.OptExp(2013),Yukawa.arXiv(2013),Lee.Science(2011),Takeda.Nature(2013)}.
On the negative side, however, since those events when a photon is heralded are totally random and uncontrollable, the probability for creating many such heralded photons simultaneously, as required for instance in linear-optical quantum computation \cite{Knill.Nature(2001),Kok.RMP(2007)}, drops exponentially with the number of photons --- unless efficient quantum memories are available.

\begin{figure*}
\centering
\includegraphics[scale=1.0,clip]{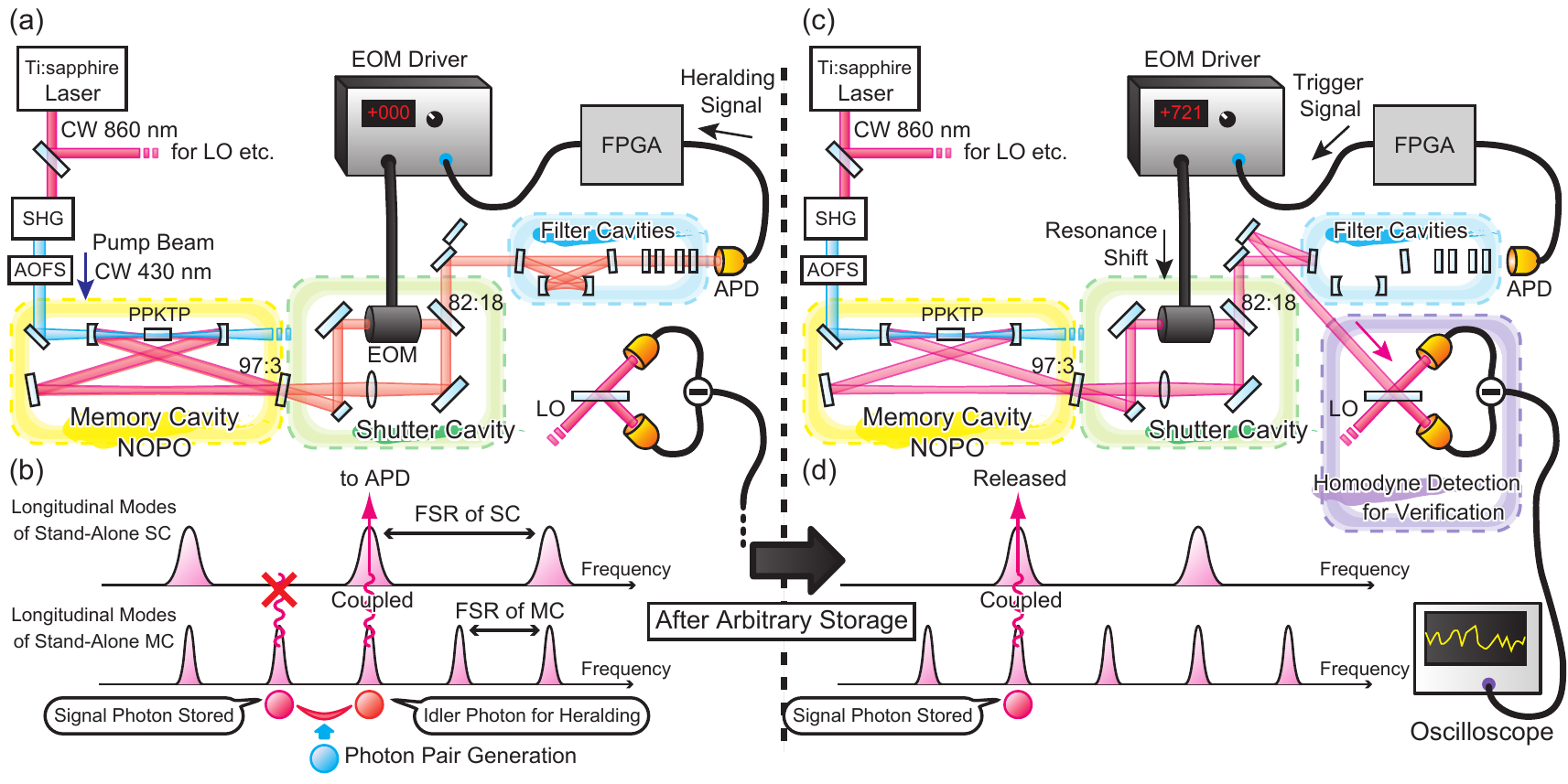}
\caption{
Schematic diagrams of storage and release.
(a) Experimental setup and beam paths at the storage stage.
(b) Conceptual diagram in the frequency domain at the storage stage.
(c) Experimental setup and beam paths at the release stage.
(d) Conceptual diagram in the frequency domain at the release stage.
Ti:Sa, titanium-sapphire laser;
CW, continuous wave;
SHG, second harmonic generator;
AOFS, acousto-optic frequency shifter;
NOPO, non-degenerate optical parametric oscillator;
PPKTP, periodically-poled KTiOPO$_4$ crystal for parametric down conversion;
EOM, electro-optic modulator;
APD, silicon avalanche photodiode;
LO, optical local oscillator;
FPGA, field-programmable gate array to process logic signals;
MC, memory cavity;
SC, shutter cavity;
FSR, free spectral range.
}\label{fig:setup}
\end{figure*}

Therefore, a possible solution to overcome the intrinsic randomness of the heralded schemes is to store a heralded state until it is required \cite{Simon.EPJD(2010)}.
Indeed, single photons were stored in matter systems in previous experiments \cite{Chaneliere.Nature(2005),Eisaman.Nature(2005),Specht.Nature(2011)}.
However, by introducing such buffer memories, some of the disadvantages of matter systems again emerge, such as low photon purity and limited tunability of the optical wavelength (though the memory efficiencies are actually improving \cite{Hedges.Nature(2010)}).
As a remedy, several all-optical experiments have been reported, however, none of them demonstrated high purity \cite{Lvovsky.NatPhoton(2009),Pittman.PRA(2002),Xu.NatPhys(2007),Tanaka.NatMater(2007),Elshaari.OE(2010)}.
Moreover, each had their own additional disadvantages: when the optical delay lines were combined with optical switches, only discrete delays were possible \cite{Pittman.PRA(2002)}; when the dynamical control of the Q factor of a photonic-crystal nanocavity was successful, the memory time scale was still only sub-nanoseconds \cite{Xu.NatPhys(2007),Tanaka.NatMater(2007),Elshaari.OE(2010)}.
In any case, it appears generally inefficient to first create flying photons and then couple them with stationary memories.
In fact, using atomic ensemble memories, Duan et al.\ proposed a scalable quantum repeater with linear optics by smartly unifying the heralding and storage (write-in) processes \cite{Duan.Nature(2001),Sangouard.RMP(2011)}.
Our scheme relies upon a similar one-step mechanism, however, in an all-optical fashion, where a heralded photon is created and automatically stored in an optical cavity.

Here, we propose the storage and controlled release of heralded quantum states by concatenating two cavities, and we experimentally demonstrate our scheme for the case of heralded single photons.
We succeeded for the first time in acquiring, effectively on demand, a quantum feature as strong as the negativity of the Wigner function.
This feature is essentially different from the more conventional single-photon character of anti-bunching \cite{Eisaman.RSI(2011)}, which is immune to linear optical losses and hence valid even for sources with very low purity.
The Wigner-function negativity rules out any description in terms of a classical phase-space distribution, and thus is important in quantum computing \cite{Mari.PRL(2012)}.
Moreover, we stress that our system is much more general
than a simple photon gun. It is rather an all-optical unification of a (potentially continuous-variable, high-dimensional) entanglement generator and a quantum memory.
In fact, our heralding mechanism can produce quantum states of light more general than single photons \cite{Bimbard.NatPhoton(2010),Yukawa.OptExp(2013),Yukawa.arXiv(2013)}, and all such states can be potentially stored in our system.
In principle, our system may function as a universal quantum memory, where optical, flying quantum information of arbitrary dimension is written into the memory via unconditional continuous-variable quantum teleportation \cite{Lee.Science(2011),Takeda.Nature(2013)}, and later recalled on demand.

\section{WORKING PRINCIPLE}

A schematic of our experimental setup with a diagram in the optical frequency domain is shown in Fig.~\ref{fig:setup}.
Inside a nondegenerate optical parametric oscillator (NOPO) operating far below threshold, signal and idler photons are probabilistically created and simultaneously appearing in different longitudinal modes.
We placed a shutter cavity (SC) at the exit of the NOPO\null.
Then, a photon inside the NOPO passes through the SC only when it is resonant with the SC, while otherwise it remains inside the NOPO\null.
Therefore, the NOPO concurrently works as a memory cavity (MC).
The switch for releasing a photon is quickly realized by shifting the resonance frequency of the SC using an electro-optic modulator (EOM).
At first, we set the SC resonant to an idler photon.
Any idler photon immediately escapes from the concatenated cavities and is then sent to a photon detector to herald the presence of its partner signal photon inside the MC [Fig.~\ref{fig:setup}(a,b)].
After the photon creation succeeded, the signal photon can be released by switching the SC whenever wanted [Fig.~\ref{fig:setup}(c,d)].
Note that an extension of this technique to multi-photon, phase-sensitive states appears to be remarkably straightforward \cite{Bimbard.NatPhoton(2010),Yukawa.OptExp(2013),Yukawa.arXiv(2013)}.
Multiple clicks of the photon detector would project the intra-cavity signal state onto a multi-photon state, and, unlike previous demonstrations without memory \cite{Bimbard.NatPhoton(2010),Yukawa.OptExp(2013),Yukawa.arXiv(2013)}, these clicks would not have to occur simultaneously.
Further note that the switching EOM, which is often lossy, is placed outside the MC\null.
This is another reason for the high purity here compared to a previous scheme based on an optical delay loop that contained the switching EOM inside the loop \cite{Pittman.PRA(2002)}.

On a more fundamental level, that intuitive explanation for the high output purity relies on the physics of general coupled systems.
More specifically, we may define the damping ratio of the coupled cavities to determine whether they are over-damped or under-damped, which is eventually reflected by the pulse shape of the released single photons.
Although this observation is certainly of great interest, it is not directly related to the main experimental achievement to be reported here, which is rather the controlled timing for releasing high-purity single photons. Therefore, we choose to leave the above issue for future works.

The experimental parameters used in our demonstration are described in the following section and in the Supplemental Material \cite{SuppMaterial}.

\section{EXPERIMENTAL RESULTS AND METHODS}
\label{sec:experiment}

\begin{figure*}
\centering
\includegraphics[scale=1.0,clip]{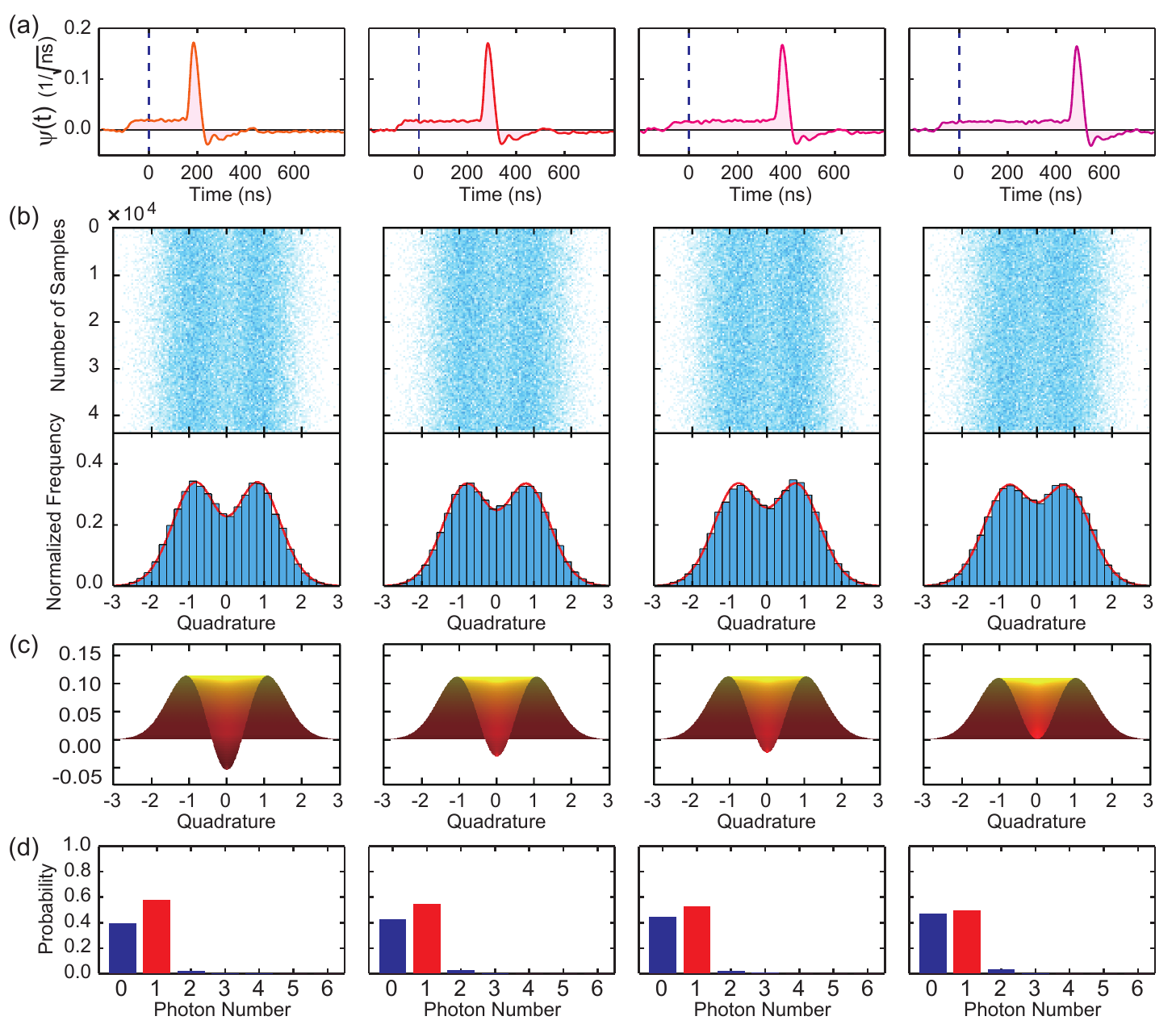}
\caption{
Experimental results for various storage times.
(a) Estimated wavepacket envelopes of the released photons.
The idler-photon detection event corresponds to $0$ ns.
(b) Samples of quadrature amplitudes (upper boxes), their normalized histograms (blue in lower boxes), and their fitting curves (red in lower boxes).
The sample number is $4.3\times10^4$.
(c) Sectional side view of the Wigner functions cutting through the phase-space origin.
(d) Photon-number distributions.
(c) and (d) correspond to the fitting curves in (b).
No correction of losses was performed in the data processing.
The storage times are $0$, $100$, $200$, and $300$ ns, in addition to an intrinsic delay of $150$ ns, corresponding to the columns from left to right.
The values of the Wigner functions at the origin are $-0.054$, $-0.030$, $-0.024$, and $0.001$ from left to right, respectively.
}\label{fig:results}
\end{figure*}

The output quantum states were characterized by optical homodyne detection.
Homodyne detection employs interference with a strong local-oscillator light beam, and hence it indicates the photons' coherence and their tunability to some specific wavelength.
This is in strong contrast to single photons verified by anti-bunching, for which two single-photon creations are necessary and quantum coherence is confirmed by the Hong-Ou-Mandel effect \cite{Sanaka.PRL(2009)}.

The experimental results are shown in Fig.~\ref{fig:results}, arranged in an array of rows and columns.
The storage time was set to $0$~ns, $100$~ns, $200$~ns, and $300$~ns in addition to the intrinsic delay of about $150$~ns after occurrence of the heralding signal, corresponding to columns from left to right.

First, we focus on the leftmost column, corresponding to the intrinsic delay.
Figure~\ref{fig:results}(a) shows the temporal shape of the wavepacket envelope $\psi(t)$ of the released flying photons, which was estimated from the raw data of homodyne detection \cite{Abdi.WIREsCS(2010)} (details can be found in the Supplemental Material).
The horizontal axis is a relative time, where $0$~ns corresponds to the events of the idler photon detection.
After opening the shutter at about $150$~ns, the wavepacket amplitude suddenly increases and rapidly goes to zero, which represents the emission of the photon.
The photon pulse width is about $50$~ns.
In fact, it can be inferred from the experimental parameters that the photon wavepacket exhibits damped oscillations (corresponding to under-damping).
This is experimentally indicated by the overshoot of the temporal shape down to negative values.
Before the photon emission, there are already nonzero values from around $-50$~ns, showing pre-leakage.
In Fig.~\ref{fig:results}(b), the quadrature samples and their histogram corresponding to the envelope function $\psi(t)$ are presented.
There is a dip at the center of the histogram, which is a characteristic of a single-photon state \cite{Lvovsky.PRA(2001),Neergaard-Nielsen.OptExp(2007)}.
By performing maximum likelihood estimation on the quadrature distribution \cite{Banaszek.PRA(1998)}, we obtained the Wigner function [Fig.~\ref{fig:results}(c)] and the photon-number distribution [Fig.~\ref{fig:results}(d)], where a dip with a negative value of $-0.054$ for the Wigner function occurs at the phase-space origin as a result of a $58.2\%$ fraction for the single-photon component.

Next, we look at the other panels to see whether the photon wavepacket alters its shape and position for longer storages.
As shown in Fig.~\ref{fig:results}(a), the emission times are correctly shifted.
However, the shape of the wavepacket is independent of the storage time, except for the pre-leakage.
These facts together indicate the success of our demonstration.
The single-photon components were estimated as $54.6\%$, $53.1\%$, and $49.7\%$, for $100$~ns, $200$~ns, and $300$~ns, respectively.
Although longer storage renders the Wigner function positive, the single photons may still be useful depending on applications owing to small multi-photon components \cite{Filip.PRL(2011)}.

\begin{figure}
\centering
\includegraphics[scale=1.0,clip]{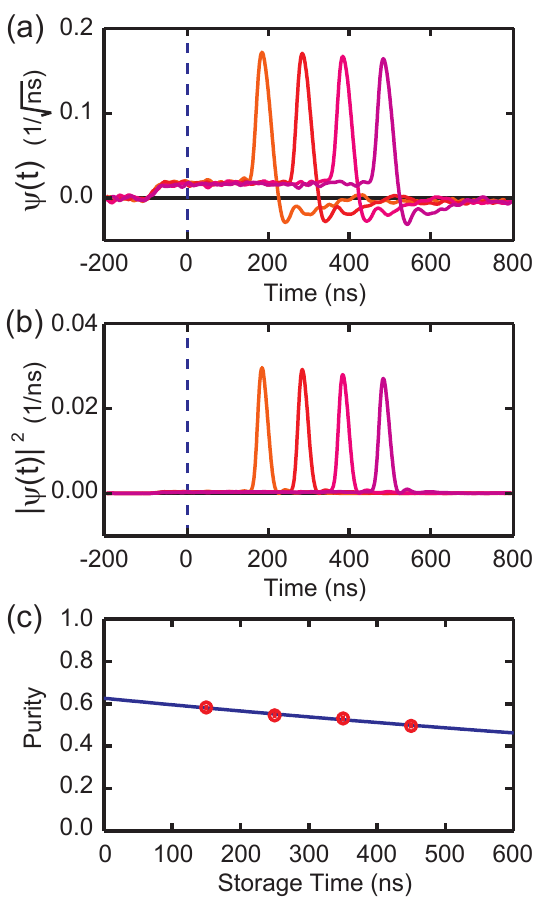}
\caption{
Experimental dependencies on the storage time.
(a) Estimated wavepacket envelopes of the released photons.
The storage times are $0$, $100$, $200$, and $300$ ns, in addition to the intrinsic $150$ ns.
(b) Absolute square of the envelopes in {\bf a}, proportional to the photon probability density with respect to time.
(c) Decay of the single-photon component with respect to the storage time, including the intrinsic delay of $150$ ns.
Red, experimental values of $58.2\%$, $54.6\%$, $53.1\%$, and $49.7\%$, with error bars of $\pm0.5\%$.
The error was roughly estimated by the bootstrap method \cite{Efron.Book(1994)}.
Blue, exponential fitting curve $P(t)=P(0)\exp(-t/\tau)$, where $P(0)=62.6\%$ and $\tau = 1.98$ $\mu$s.
}\label{fig:results_summary}
\end{figure}

All our results are combined in Fig.~\ref{fig:results_summary} for an easy comparison between the envelope function $\psi(t)$ [Fig.~\ref{fig:results_summary}(a)], its absolute square $|\psi(t)|^2$ [Fig.~\ref{fig:results_summary}(b)], and the single-photon component [Fig.~\ref{fig:results_summary}(c)].
The probability density for the presence of a photon at time $t$ is proportional to the absolute square $|\psi(t)|^2$.
From Fig.~\ref{fig:results_summary}(b), we can infer that the contribution of the pre-leakage is very small.
The lifetime of the photon storage is of the order of $1$~$\mu$s, as can be seen from the fitting curve in Fig.~\ref{fig:results_summary}(c), which is comparable with the lifetimes of some ensemble memories \cite{Hedges.Nature(2010)}.
In principle, this lifetime can be further extended by decreasing the intracavity losses.
In the following, we will now summarize our experimental methods.

\subsection{Procedure}

Our setup is shown in Fig.~\ref{fig:setup}.
The source laser is a continuous-wave Ti:sapphire laser with a wavelength of $860$~nm.
A part of the laser output is frequency-doubled to $430$~nm, which is used as a pump beam for the NOPO.
Two longitudinal modes of the NOPO at $860$~nm, separated by a free spectrum range (FSR), are used as the signal and idler.
A single longitudinal mode (idler) is selected by filter cavities, and a photon detection projects the twin mode (signal) onto a single-photon state.
After some predetermined waiting time, a high voltage of $721$~V is applied to the EOM to release the signal photon.
It was experimentally essential to avoid above-threshold oscillation at the unused longitudinal modes of the NOPO\null.
The photon generation rate was about $300$ per second with a NOPO pumping power of $3$~mW\null.
The power of the LO was $18$~mW, which produced an optical shot noise about $20$~dB above the detector's electronic noise.

\subsection{Cavity Configuration}

The bow-tie shaped MC contains a periodically-poled KTiOPO$_4$ (PPKTP) crystal as a nonlinear medium to operate as a NOPO; it has a round-trip length of $1.4$~m (corresponding to the FSR of $2.2\times10^2$~MHz), where the round-trip optical loss was $0.2$--$0.3\%$.
These values determine the photon lifetime.
The SC contains a RbTiOPO$_4$ (RTP) EOM for the resonant-frequency shift; it has a round-trip length of $0.7$~m, where the round-trip optical loss was $3\%$.
The coupler reflectivity between the MC and the SC is $97\%$, while that between the SC and the external field is $83\%$.

\subsection{Verification}

While the homodyne detection is a phase-sensitive measurement, the single-photon state is a phase-insensitive state.
This phase-insensitivity was confirmed by the invariance of the quadrature distribution with the scanned and unscanned phase of the local oscillator.
The photon envelope $\psi(t)$ is determined by the principal component analysis (PCA) \cite{Abdi.WIREsCS(2010)}.
Each quadrature sample is obtained by a weighted integral of the continuous homodyne data $x(t)$ as $\int x(t)\psi(t)dt$.
When we replaced the idler beam by other, uncorrelated light before the photon detection, the observed signal state was almost a zero-photon state, which ensures that the original signal field was indeed conditionally obtained via the idler detection.

Further information on the experimental methods can be found in the Supplemental Material \cite{SuppMaterial}.

\section{CONCLUSION}

In conclusion, we proposed and experimentally demonstrated the insertion of a quickly tunable cavity at the exit of a NOPO, effectively serving as a kind of quantum memory.
After a heralded creation, storage, and on-demand release of single photons with this system, the output states exhibited strong nonclassicality, as became manifest through their negative Wigner functions.
Our device works at normal temperature and pressure.
Moreover, our scheme is compatible with hybrid quantum optical experiments, where photon and homodyne detections are employed at the same time, like quantum teleportation \cite{Lee.Science(2011),Takeda.Nature(2013)} and entanglement distillation \cite{Takahashi.NatPhoton(2010),Fiurasek.PRA(2010)}.
Many advanced quantum demonstrations are possible with our system in the future, e.g., the storage and release of multi-photon and/or phase-sensitive states of light \cite{Bimbard.NatPhoton(2010),Yukawa.OptExp(2013),Yukawa.arXiv(2013)}, the controlled interference of many photons simultaneously released from independent memories \cite{Knill.Nature(2001),Kok.RMP(2007),Sanaka.PRL(2009)}, and the deterministic transfer of arbitrary quantum optical states into the memories via unconditional continuous-variable quantum teleportation \cite{Lee.Science(2011),Takeda.Nature(2013)}.

\begin{acknowledgments}
This work was partly supported by 
PDIS, GIA, G-COE, and APSA commissioned by the MEXT of Japan,
FIRST initiated by the CSTP of Japan, 
and REFOST of Japan. 
P.v.L.\ acknowledges support from QuOReP of the BMBF in Germany.
\end{acknowledgments}

\clearpage

\begin{center}
\noindent
{\bf
Supplemental Material for \\
Creation, Storage, and On-Demand Release \\
of Optical Quantum States \\
with a Negative Wigner Function}
\end{center}

\section{PRE-LEAKAGE DUE TO IMPERFECTION OF THE SHUTTER DECOUPLING}

In the analysis based on the principal component analysis (PCA) in the main text, the pre-leakage of the signal photon also contributes to the purity.
The obtained wavepackets are informative in the sense that they show the actual shape of the photon wavepacket, however, the inclusion of the pre-leakage to the analysis reveals an overestimation of the memory lifetime.
An important figure here is the probability of a single photon in a mode time-shifted in accordance with the storage time, because it contributes to a successful multi-photon interference from independent memories.

\begin{figure*}[tbp]
\centering
\includegraphics[scale=1.0,clip]{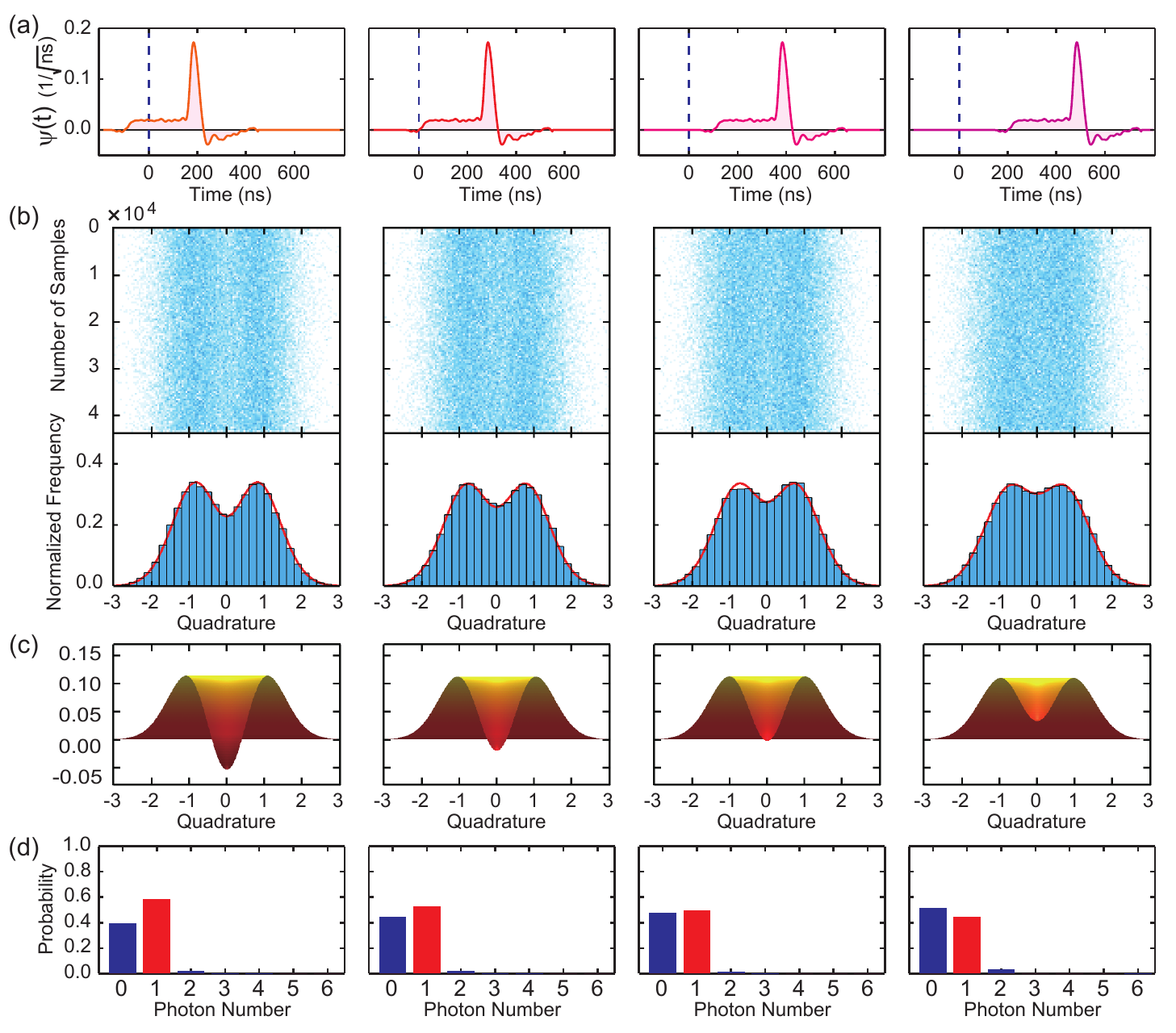}
\caption{
Experimental results for various storage times with time-shifted wavepackets.
(a) Wavepacket envelopes.
The idler-photon detection corresponds to $0$ ns.
(b) Samples of quadrature amplitudes (upper boxes), their normalized histograms (blue in lower boxes), and their fitting curves (red in lower boxes).
(c) Sectional side view of the Wigner functions cutting through the phase-space origin.
(d) Photon-number distributions.
(c) and (d) correspond to the fitting curves in (b).
The storage times are $0$, $100$, $200$, and $300$ ns, in addition to an intrinsic delay of $150$ ns, corresponding to the columns from left to right.
The values of the Wigner functions at the origin are $-0.054$, $-0.020$, $-0.002$, and $0.033$ from left to right, respectively.
}\label{fig:results_shifted}
\end{figure*}

\begin{figure}[tbp]
\centering
\includegraphics[scale=1.0,clip]{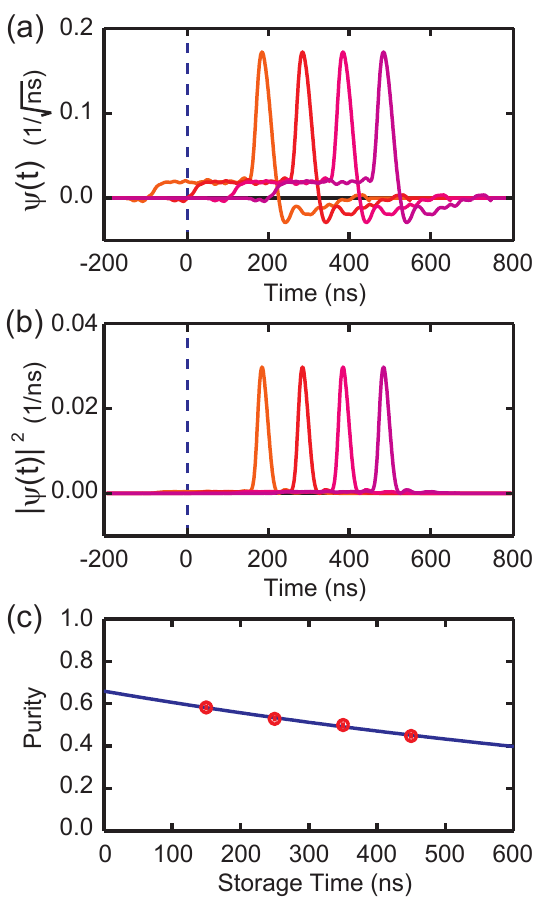}
\caption{
Experimental dependencies on the storage time with time-shifted wavepackets.
(a) Wavepacket envelopes.
The storage times are $0$, $100$, $200$, and $300$ ns, in addition to the intrinsic $150$ ns.
(b) Absolute square of the envelopes in (a), proportional to the photon probability density with respect to time.
(c) Decay of the single-photon component with respect to the storage time, including the intrinsic delay of $150$ ns.
Red, experimental values of $58.2\%$, $52.9\%$, $49.9\%$, and $44.8\%$, with error bars of $\pm0.5\%$.
Blue, exponential fitting curve $P(t)=P(0)\exp(-t/\tau)$, where $P(0)=65.9\%$ and $\tau = 1.19$ $\mu$s.
}\label{fig:results_summary_shifted}
\end{figure}

In Figs.~\ref{fig:results_shifted} and \ref{fig:results_summary_shifted}, we show the results for time-shifted wavepackets.
The wavepacket estimated from the results of the intrinsic delay is the original one.
Here, the time bin from $-150$~ns to $450$~ns was clipped, and the wavepacket then renormalized.
Subsequently, it was shifted by $100$~ns, $200$~ns, and $300$~ns to analyze the correspondingly delayed results.
No correction of losses was performed in the data processing.

We anticipate a reduction of the pre-leakage by increasing the number of the shutters serially.
Another interesting option would be to monitor the pre-leakage by a photon detector.
If we detect a photon before opening the shutter, we can simply discard the stored light and start a new creation of quantum states.
In this way, the purity will be increased compared to the case of not monitoring the pre-leakage.

\section{SUPPLEMENTAL METHODS}

In this section, we will describe the experimental methods and parameters, which are not included in the main text, except for the quantum state reconstruction from the homodyne outcomes, which will be discussed separately in the next section.

The schematic setup is shown in Fig.~\ref{fig:setup} of the main text.
The contents of this section can be listed as follows:
first, chopping of bright beams,
second, filtering of the idler path,
third, detuning of optical frequencies,
fourth, data storage by the oscilloscope,
fifth, other experimental parameters.

\subsection{Feedback Control of the Optical System and Beam Chopping}

In order to keep the cavities resonant to the laser beams, we utilized interference of auxiliary bright beams, from which we obtained error signals for classical feedback control.
However, such bright beams, except for the local oscillator (LO) beam and the pump beam, can induce fake clicks or even break down the avalanche photodiode (APD).
Therefore, in the experiment, we switched the optical system cyclically between two phases:
one is the phase for actively controlling the optical system with classical feedback by using bright beams (classical phase), and the other is the phase for manipulating the quantumness at the single-photon level by blocking the bright beams (quantum phase).

The blocking and unblocking of the bright beams was achieved by switching on and off the diffraction via two acousto-optic modulators (AOMs).
One AOM upshifted the frequency, while the other AOM downshifted it to cancel out the frequency shift.
The cavity lengths, which determine the cavity resonances, were controlled by using piezoelectric transducers (PZTs), except for the fast switching of the shutter cavity (SC), which was achieved by an electro-optic modulator (EOM).
During the quantum phase, the resonance of the cavities was held by keeping the voltages applied to the PZTs.
We also placed a switching AOM just before the avalanche photodiode (APD), so that only the beam at the quantum phase can enter the APD.
Note that these AOMs and PZTs, as well as several EOMs that modulated the bright beams for the Pound-Drever-Holl locking of the cavities, are omitted from Fig.~\ref{fig:setup} of the main text.

We repeated the quantum-classical phase changes at a rate of $5$~kHz and with a time ratio of $2:3$.
Thus, the $300$ counts per second for the single-photon generation in the experiment effectively
correspond to $750$ counts per second if we could maintain the quantum phase.
The repetition was controlled by the same FPGA that controlled the storage time after the heralding signals.
The heralding signals were gated by the FPGA, so that only those at the quantum phase drive the switching EOM of the SC.

\subsection{The Idler Beam Path and Filter Cavities}

We utilized three filter cavities.
One was a bow-tie shaped ring cavity with a round-trip length of $250$~mm, and the other two were Fabry-Perot cavities with a round-trip lengths of several~mm.
The first ring cavity also worked as a spatial separator of the signal and idler beams.
These three cavities had different free spectral ranges (FSRs) to prevent beams other than the target idler from entering the APD.
Between the two Fabry-Perot cavities, we inserted an optical isolator in order to decouple the two cavities.
After the final filter cavity, as noted above, we placed an AOM for switching the beam path.
A spatial single mode was selected by a single-mode fiber before the APD to reduce fake clicks by stray light.

The transmission efficiencies of the filter cavities were $90\%$, $76\%$, and $65\%$.
That of the optical isolator was $88\%$.
The diffraction efficiency of the AOM was $66\%$.
The coupling efficiency to the single-mode fiber was $83\%$.
The APD was an SPCM-AQRH-16 (Excelitas Technologies), whose dark counts were $5$ counts per second (taking into account the time ratio, $13$ counts per second).
Note that the losses on the idler line do not deteriorate the purity of the heralded signal-photon states, but only decrease the photon generation rate, including a relative increase of the dark-count contribution.

\subsection{Optical Frequencies}

The bright light source was a continuous wave (CW) Ti:sapphire laser operating at $860$~nm, whose frequency is denoted by $\omega_\text{source}$.
The optical frequency of the pump, signal, and idler beams for the nondegenerate optical parametric oscillator (NOPO) are denoted by $\omega_\text{pump}$, $\omega_\text{signal}$, and $\omega_\text{idler}$, respectively.
From the energy conservation condition of the parametric down conversion, we have $\omega_\text{pump} \approx \omega_\text{signal} + \omega_\text{idler}$.
Here, since we must take into account the finite bandwidth of each beam, we use the approximately equal sign `$\approx$' instead of the equal sign `$=$'.
For the characterization of the emitted light wavepacket, we employed the output of the Ti:sapphire laser without frequency shift as the LO for the homodyne detection, thus $\omega_\text{signal} \approx \omega_\text{LO} = \omega_\text{source}$.
The signal and idler frequencies were separated by nearly the FSR of the NOPO $\Delta\omega_\text{NOPO}$, thus $\omega_\text{idler} \approx \omega_\text{source} \pm \Delta\omega_\text{NOPO}$, while we did not distinguish plus or minus in the experiment.
Therefore, the pump beam was the second harmonic of the light source with a frequency shift about the FSR, $\omega_\text{pump} \approx 2\omega_\text{source} \pm \Delta\omega_\text{NOPO}$.
The frequency of the second harmonic was shifted by an acousto-optic frequency shifter (AOFS) to produce the pump beam, as depicted in Fig.~\ref{fig:setup} of the main text.

However, as will be discussed below, we made use of detuning for various experimental reasons.
The pump frequency shift $\omega_\text{pump} - 2\omega_\text{source}$ was not exactly $\Delta\omega_\text{NOPO}$, but detuned by $5$~MHz.
In addition, the signal frequency $\omega_\text{signal}$ was not exactly at the LO frequency $\omega_\text{source}$, but detuned by $300$--$500$~kHz.

The former detuning was applied on the pump beam by the AOFS.
The latter detuning corresponds to a shift of the resonance of the NOPO.
However, this detuning was also realized via an AOFS, which is omitted from Fig.~\ref{fig:setup} of the main text.
It was achieved by shifting the frequency of the auxiliary bright beam
utilized for the locking of the NOPO.

Next, we explain the experimental reasons for the two detunings.

\subsubsection{Suppressing above-threshold oscillation}

As mentioned in Sec.~\ref{sec:experiment} of the main text, our NOPO starts above-threshold oscillation at unused longitudinal modes unless detuning is utilized.
Since the longitudinal modes are almost equally spaced by the FSR $\Delta\omega_\text{NOPO}$ in the frequency domain, whenever a pair of $\omega_\text{signal}$ and $\omega_\text{idler}$ satisfies the energy conservation condition, there are other pairs of longitudinal modes $\omega_\text{signal}+n\Delta\omega_\text{NOPO}$ and $\omega_\text{idler}-n\Delta\omega_\text{NOPO}$ with an integer $n$ that also satisfy the energy conservation condition.
Every longitudinal mode has a different finesse due to the frequency dependence of the SC.
The higher the finesse is, the stronger is the intracavity vacuum field of the mode.
As a result, every pair that satisfies the energy conservation has a different gain of the down conversion, proportional to the product of the corresponding intra-cavity field strengths.
In the experiment, we used the pair of a high-finesse mode (signal) and a low-finesse mode (idler) at the memory excitation (write-in) stage.
However, almost inevitably, pairs exist that are composed of two high-finesse modes.
Such pairs have much larger gain of the down conversion than the selected signal and idler pair, thus the NOPO easily goes beyond threshold for these pairs.

On the other hand, the high-finesse modes have much narrower bandwidth than the low-finesse modes.
Thus, in principle, the down conversion gain becomes very small with a small detuning of the pump frequency for a pair of two high-finesse modes, while it remains almost unchanged for a pair of high- and low-finesse modes.
By this technique, the above-threshold oscillation can be avoided.

Experimentally, a relatively large detuning of $5$~MHz was introduced to remove any above-threshold oscillation.
Another option would be to insert wavelength-selective components inside the MC or the SC. However, since such components are typically lossy, we did not explore this option in our demonstration.

\subsubsection{Reducing storage of scattered light}

In the experiment, the LO beam for the homodyne detection was not switched.
Therefore, the LO was being scattered during the quantum phase, mainly at the photodiodes of the homodyne detector.
We observed that the scattered LO, corresponding to a weak coherent state at the single-photon level, was stored inside the MC and induced phase-space displacements of the stored signal states.
This can be avoided though, with only a small deterioration of the measured quality of the single-photon states, by slightly detuning the signal-mode frequency of the MC from that of the LO.
We detuned the signal frequency by $300$~kHz--$500$~kHz, which is much smaller than the released photon bandwidth of about $20$~MHz and thus did not significantly degrade the photon quality observed by the homodyne detection.
The detuning was achieved by a small shift of the MC round-trip length from the LO resonance, controlled by a PZT.

We did not take much care of the exact value of the detuning, because the target quantum state was a phase-insensitive single-photon state.
If we did want to write in a phase-sensitive state, since the stored state during its storage is expected to rotate in phase space relative to the LO frame at the detuning frequency, we would have to either compensate this rotation or remove the displacements from the LO scattering according to the detuning condition.
However, we stress that the scattered light in our experiment was solely a technical problem of the verification part of the setup, and not at all a fundamental deficiency of the memory system itself.

\subsection{Data Storage}

The heralding signals, which were processed by the FPGA, were not only to trigger the EOM driver, but also to trigger the oscilloscope storage of the homodyne signal.
The oscilloscope captured the homodyne signal for $1$~$\mu$s around each trigger signal.
The resolution of the analog-to-digital conversion was $8$~bits and the sampling rate was $1$~GS/s.
We repeatedly acquired the frame $4.3\times10^4$ times for each condition.

\subsection{Other Parameters}

In addition to the parameters given above and in Sec.~\ref{sec:experiment} of the main text,
there are the following experimental parameters:
the propagation efficiency from the exit of the SC to the homodyne detector was $97\%$,
the quantum efficiency of the homodyne detector was $99\%$,
and the interference visibility between the LO and the output beam of the SC was $99\%$.

The size of the second-order nonlinear optical PPKTP crystals for the NOPO and the SHG was $10$~mm in length and $1$~mm by $1$~mm in cross section.
The radius of curvature of the two concave mirrors of the NOPO was $150$~mm, while the focal length of the lens inside the SC was $500$~mm.

The FPGA that controled the bright beam chopping and the delay of the trigger signal to shift the resonance of the SC was Virtex-4 (Xilinx).
The FPGA clock period was $10$~ns, and thus, in our system, the photon emission times were discretized into $10$~ns intervals, whereas the optical system itself would, of course, allow emissions continuous in time.

\section{VERIFICATION BY HOMODYNE DETECTION}

As described above, we recorded a $1$~$\mu$s frame of homodyne signal around each heralding signal.
Every frame contained a single measurement result corresponding to a single-photon state.
However, every frame also contained vacuum noise of other orthogonal modes.
We must therefore extract the information of the target single photon from such background noise,
which is a big difference from click-by-click photon detection.
In this section, we will compare the homodyne and photon detections, and discuss how the information of the single-photon state can be extracted from the homodyne signal.

The contents of this section are as follows.
We will first review the mathematical description of the output wavepacket, introducing the mode function.
The mode function is equivalent to the wavepacket envelope used in the main text, however, here we choose a different terminology.
Next, we will describe both the photon and the homodyne detection, with a focus on optical coherence.
Then, we explain how to estimate the mode function via the PCA.
Finally, we briefly describe how to represent quantum states by Wigner functions.

\subsection{Mathematical Description of a Flying Single-Photon State}

Since the transverse modes and the polarization are determined by the LO, here we focus on the longitudinal modes of the wavepackets.
We use the time $t$ as a continuous degree of freedom in the longitudinal direction, which could also be, alternatively, the spatial coordinate $z$, with $z=ct$ and $c$ the speed of light.
We define annihilation and creation operators at time $t$ as $\hat{a}(t)$ and $\hat{a}^\dagger(t)$, respectively.
Their commutation relation is,
\begin{align}
[\hat{a}(t),\hat{a}^\dagger(t^\prime)] = & \delta(t-t^\prime), &
[\hat{a}(t),\hat{a}(t^\prime)] = & 0,
\end{align}
where $\delta(t)$ is the Dirac delta function.
The photon density operator at time $t$ is defined as,
\begin{align}
\hat{n}(t)\equiv\hat{a}^\dagger(t)\hat{a}(t).
\end{align}

Similarly, we define annihilation and creation operators for a wavepacket mode specified by a mode function $\psi(t)$ as follows,
\begin{align}
\hat{a}_\psi^\dagger = & \int dt\psi(t)\hat{a}^\dagger(t), &
\hat{a}_\psi = & \int dt\psi^\ast(t)\hat{a}(t),
\end{align}
where the superscript $\ast$ denotes the complex conjugate.
In the following, for simplicity, we often drop the argument $(t)$ of the mode function.
The usual single-mode bosonic commutation relations are equivalent to the normalization of the mode function $(\psi,\psi)=1$, with the inner product of two mode functions,
\begin{align}
(\psi,\phi) \equiv \int dt\psi^\ast(t)\phi(t).
\end{align}
The corresponding photon-number operator is defined as,
\begin{align}
\hat{n}_{\psi} \equiv \hat{a}_{\psi}^\dagger\hat{a}_{\psi}.
\end{align}
Note that $\int dt|\psi(t)|^2\hat{n}(t)$ is very different from $\hat{n}_{\psi}$.
Related to this fact is the impossibility to determine the photon number in a given wavepacket $\psi$ in a simple way using photon detection, as will be discussed later.

An infinite set of orthonormal mode functions $\{\psi_k\}_{k\in\natnum}$ specifies orthogonal modes such that
\begin{align}
[\hat{a}_{\psi_k},\hat{a}_{\psi_\ell}^\dagger] = & \delta_{k\ell}, &
\text{iff}\quad (\psi_k,\psi_\ell) = & \delta_{k\ell},
\end{align}
where $\delta_{k\ell}$ is the Kronecker delta.
We choose an infinite number of such orthonormal functions $\{\psi_k\}_{k\in\natnum}$ to obtain a basis for the function space.
Then, the creation operator of an arbitrary mode $\psi$ is expressed in this basis as,
\begin{align}
\hat{a}_\psi^\dagger = & \sum_{k\in\natnum}(\psi_k,\psi)\hat{a}_{\psi_k}^\dagger.
\label{eq:cr_op_decomp}
\end{align}
Since the number operators $\{\hat{n}_{\psi_k}\}_{k\in\natnum}$ of orthonormal modes are mutually commutative and independent, they have simultaneous eigenstates with independent eigenvalues $n_k$,
\begin{align}
\ket{n_0\,_{\psi_0},n_1\,_{\psi_1},n_2\,_{\psi_2},\dots} = \frac{\hat{a}_{\psi_0}^{\dagger n_0}}{\sqrt{n_0!}}\frac{\hat{a}_{\psi_1}^{\dagger n_1}}{\sqrt{n_1!}}\frac{\hat{a}_{\psi_2}^{\dagger n_2}}{\sqrt{n_2!}}\dots\ket{\varnothing},
\end{align}
where $\ket{\varnothing}$ is the zero-photon (vacuum) state, $\hat{a}(t)\ket{\varnothing}=0$ for all $t$.
An arbitrary pure state is a superposition of such number states.
This also means that the overall Hilbert space $H$ of the multi-mode quantum states can be decomposed into a tensor product of single-mode Hilbert spaces,
\begin{align}
H = H^{\psi_0} \otimes H^{\psi_1} \otimes H^{\psi_2} \otimes \dots.
\end{align}
Operators acting on the total Hilbert space $H$ may also be decomposed into such tensor products, e.g., $\hat{a}_{\psi_0}^\dagger=\hat{a}^\dagger\otimes\hat{1}\otimes\hat{1}\otimes\dots$ where $\hat{1}$ is the identity operator.
However, this does not hold in general.
For example, a density operator $\hat{\varrho}$ which cannot be expressed as a convex sum of tensor products describes an entangled (inseparable) state.
The single-mode density operator $\hat{\rho}^{\psi_0}$ is obtained by tracing out all the orthogonal modes,
\begin{align}
\hat{\rho}^{\psi_0} = \Tr_{\bigotimes_{k\neq0} H^{\psi_k}}\hat{\varrho},
\end{align}
where $\Tr_{H^X}$ is the partial trace over a Hilbert space $H^X$.
In the following, we will use $\hat{\varrho}$ for an overall density operator and $\hat{\rho}$ for a density operator reduced to a single mode.

An ideal single-photon state, for which exactly one photon is present in some wavepacket $\psi$, is expressed as
\begin{align}
\ket{1_\psi} = \hat{a}_\psi^\dagger\ket{\varnothing} = \int dt\psi(t)\hat{a}^\dagger(t)\ket{\varnothing}.
\end{align}
Realistic single-photon states, however, are effectively mixtures of the above pure single-photon state and the vacuum state,
\begin{align}
\hat{\varrho}_{\psi,p}=p\ket{1_\psi}\bra{1_\psi}+(1-p)\ket{\varnothing}\bra{\varnothing}.
\label{eq:rho_expr}
\end{align}
From the decomposition in Eq.~\eqref{eq:cr_op_decomp}, the reduced density operator of the mode $\psi_0$ is,
\begin{align}
\hat{\rho}^{\psi_0}
& = \Tr_{\bigotimes_{k\neq0} H^{\psi_k}}[p\ket{1_\psi}\bra{1_\psi}+(1-p)\ket{\varnothing}\bra{\varnothing}] \notag\\
& = p|(\psi_0,\psi)|^2\ket{1}\bra{1} + [1-p|(\psi_0,\psi)|^2]\ket{0}\bra{0},
\end{align}
where $\ket{n}$ is a single-mode $n$-photon state.
As expected, the purity is maximized by setting the observed mode function $\psi_0$ equal to the ideal mode function of the single photon $\psi$.
However, whenever these two functions are different, the purity decreases depending on the mode mismatch from $p$ to $p|(\psi_0,\psi)|^2$.

For a single-photon state, the photon density with respect to time $t$ is,
\begin{align}
P(t) = \Tr(\hat{n}(t)\ket{1_\psi}\bra{1_\psi}) = |\psi(t)|^2.
\end{align}
Therefore, as mentioned in the main text, the probability density for the presence of a photon at time $t$ is proportional to the absolute square of the mode function.

\subsection{Homodyne Detection versus Photon Detection}

A photon detector, such as an APD in the Geiger mode, detects photons click by click in a particle-like manner, whereas a homodyne detector observes the wave-like interference of the signal with a LO.
Therefore, unlike photon detection, a single homodyne outcome can never show the presence of a photon.
However, from the statistics of the outcomes, we can estimate the state -- a technique known as quantum tomography.
Homodyne characterization is advantageous in several respects, such as high quantum efficiencies and the full characterization of a single optical mode. On the other hand, it is also demanding, because it requires a light source coherent with the target field.

What we would like to stress here is that the sole capability of homodyne verification is already an indication for the coherence of the obtained single photons.
This is in strong contrast to photon-detection-based verification relying upon the observation of anti-bunching. The latter negates the simultaneous presence of photons, but it contains no information about the coherence.

Another important point is that the anti-bunching is immune to linear optical losses.
On the other hand, the purity, which is given by the probability of the corresponding wavepacket to contain exactly one photon, also directly represents the success probability for creating single photons in the specific wavepackets coherently.
The purity is degraded by optical losses, and in this sense, it is generally much more difficult to obtain good purity than to obtain good anti-bunching.

\subsubsection{Photon detection}

Here we suppose that the speed of the photon detector is high enough compared to the time scale of the photon wavepacket.
Then, the observable of the photon detection at time $t$ is well approximated by the photon density operator $\hat{n}(t)\equiv\hat{a}^\dagger(t)\hat{a}(t)$.
Given a density operator $\hat{\varrho}$ containing at most one photon, the probability density of the photon detection is $P(t) = \eta\Tr(\hat{n}(t)\hat{\varrho})$, where $\eta$ is the quantum efficiency of the detection system.
Inserting the density operator $\hat{\varrho}_{\psi,p}$ from Eq.~\eqref{eq:rho_expr}, we obtain the probability density to detect a photon at time $t$,
\begin{align}
P(t) = p\eta|\psi(t)|^2.
\end{align}
Therefore, by taking statistics of the detection time $t$, we gain information about the mode function without its phase component, $|\psi(t)|$.
The photon probability $p$ is estimated from $\int dtP(t)$ if the quantum efficiency $\eta$ is known.
However, as already mentioned, detecting a photon with a probability density $\eta|\psi(t)|^2$ can never be a proof for the presence of a photon in a wavepacket $\psi$.
For example, we are not able to distinguish functions different in phase, $\psi(t)$ and $\psi(t)e^{i\theta(t)}$.
Note that an optical frequency shift by $\omega$ corresponds to setting $\theta(t)=\omega t$.
Furthermore, we cannot even distinguish a pure state from a decohered mixed state.
In order to show this, we consider smooth functions $\phi_\tau(t)\equiv\phi(t-\tau)$ in order to deconvolute the photon density as $|\psi(t)|^2=\int d\tau p(\tau)|\phi(t-\tau)|^2$, where $(\phi,\phi) = 1$ and $\int d\tau p(\tau) = 1$.
Then, a mixed state,
\begin{align}
\int d\tau p(\tau)\ket{1_{\phi_\tau}}\bra{1_{\phi_\tau}},
\label{eq:decohered_photon}
\end{align}
gives the same probability density $P(t)=\eta|\psi(t)|^2$ as a pure single-photon state $\ket{1_\psi}\bra{1_\psi}$.
However, the single-photon purity with respect to the wavepacket $\psi_0=\psi$ is decreased from $1$ to,
\begin{align}
\bra{1}\Tr_{\bigotimes_{k\neq0} H_{\psi_k}}\bigl[\int d\tau p(\tau)\ket{1_{\phi_\tau}}\bra{1_{\phi_\tau}}\bigr]\ket{1} \notag\\
= \int d\tau p(\tau)|(\phi_\tau,\psi)|^2.
\end{align}

The discrepancy between a photon click and the actual purity associated with some mode function $\psi_0$ may be explained as follows.
The quantum state $\hat{\rho}^{\psi_0}$ must be estimated from the observables on the mode $\psi_0$, constructed from $\hat{a}_{\psi_0}$ and $\hat{a}_{\psi_0}^\dagger$.
However, any observable constructed from $\hat{n}(t)$ can never provide this due to the lack of interference terms $\hat{a}^\dagger(t)\hat{a}(t^\prime)$ for $t\neq t^\prime$.
As will be discussed later, in contrast, homodyne detection can provide observables that are constructed from $\hat{a}_{\psi_0}$ and $\hat{a}_{\psi_0}^\dagger$, which allows us to estimate $\hat{\rho}^{\psi_0}$.

Since the coherence, as well as the stability of the frequency, etc., is crucial for interferometric usage, these properties should be ensured experimentally.
For this purpose, in photon detection experiments, we typically create two photons and check the beam splitter interference, by which we observe the photon bunching effect depending on their indistinguishability, i.e., whether they have identical frequencies, polarizations, and so forth; but not whether some decoherence like in Eq.~\eqref{eq:decohered_photon} occurs.

So far, we have discussed photon detection on fields which contain at most one photon.
In the experiment, also the effective vanishing of the multi-photon probability must be checked to verify the single-photon character.
This can be achieved by observing the anti-bunching effect, which appears in the second-order photon correlation,
\begin{align}
g^{(2)}(\tau) & = \frac{\langle:\hat{n}(t)\hat{n}(t+\tau):\rangle}{\langle\hat{n}(t)\rangle\langle\hat{n}(t+\tau)\rangle} \notag\\
& = \frac{\langle\hat{a}^\dagger(t)\hat{a}^\dagger(t+\tau)\hat{a}(t+\tau)\hat{a}(t)\rangle}{\langle\hat{a}^\dagger(t)\hat{a}(t)\rangle\langle\hat{a}^\dagger(t+\tau)\hat{a}(t+\tau)\rangle},
\end{align}
where $\langle\hat{X}\rangle$ is the expectation value of an observable $\hat{X}$, and $:\hat{X}:$ is the normally ordered form of an operator $\hat{X}$.
It is obtained by using two photon detectors (the Hanbury-Brown and Twiss setup).
If the multi-photon probability is zero at any instant of time, we have $g^{(2)}(0)=0$.
In contrast, when the photon probabilities are totally uncorrelated, $g^{(2)}(\tau)=1$, which is typically the case for $\tau\to\infty$.
Therefore, the anti-bunching is typically observed as a gap at $\tau=0$ in the correlation function $g^{(2)}(\tau)$.
A coherent state shows a Poissonian statistics corresponding to $g^{(2)}(0)=1$, irrespective of its average photon number, and we have $g^{(2)}(0)\ge1$ for any incoherent mixture of coherent states.
Therefore, a measure of the quality of a single-photon stream is, beyond the classical limit of one, how much $g^{(2)}(0)$ gets closer to zero.
Note that $g^{(2)}(\tau)$ itself says nothing about the coherence of the single photons, similar to the case of a single-photon detector discussed above.

Another point is that $g^{(2)}(0)$ is the multi-photon probability normalized by the single-photon probabilities, such that it quantifies the deviation from a random photon emission.
Therefore, $g^{(2)}(\tau)$ remains unchanged under linear optical losses
(i.e., the effect of $\hat{a}(t)\to\sqrt{\eta}\hat{a}(t)$ is canceled between the numerator and denominator).
This is both good and bad.
On the one hand, it can verify the nonclassicality of a very faint light beam, where the photon detection events only scarcely occur.
On the other hand, it even validates very inefficient single-photon sources, even though these may be actually creating wavepackets very close to vacuum states.
Such inefficient, rarely successful single-photon generators are not useful for many applications.

On the other hand, the single-photon purity measures how much a given wavepacket $\psi_0$ contains the ideal single-photon state in a coherent manner.
Especially, when its value exceeds one half, a stronger form of nonclassicality, that is the negativity of the Wigner function, emerges.
The density matrix and the Wigner function are usually obtained by homodyne tomography, as we discuss next.

\subsubsection{Homodyne detection}

The annihilation and creation operators $\hat{a}$ and $\hat{a}^\dagger$ in the particle picture correspond to complex and complex conjugate amplitudes of the light field in the wave picture.
They are not observables, because they are not self-adjoint.
However, the quadrature amplitudes are self-adjoint,
\begin{align}
\hat{x} \equiv & \frac{\hat{a}+\hat{a}^\dagger}{\sqrt{2}}, &
\hat{p} \equiv & \frac{\hat{a}-\hat{a}^\dagger}{i\sqrt{2}}.
\end{align}
Analogously to the quantum mechanical position and momentum operators, $\hat{x}$ and $\hat{p}$ cannot be simultaneously measured with perfect precision, since they do not commute with each other, $[\hat{x},\hat{p}]=i$.
The instantaneous outcome of the homodyne detection is, for an ideal detector with infinite bandwidth, a quadrature at time $t$, $\hat{x}(t)=[\hat{a}(t)+\hat{a}^\dagger(t)]/\sqrt{2}$, as shown below.
Note that realistically, due to the finite bandwidth of the detector, we can never measure a truly instantaneous quadrature, which would have an infinitely large variance.
However, since we are interested in a situation where the bandwidth of single photons is much narrower than that of the homodyne detector, we may faithfully equate an instantaneous homodyne outcome with an instantaneous quadrature outcome.

Homodyne detection measures the interference between a LO beam and a signal light field.
The LO has a large coherent amplitude $\alpha = |\alpha|e^{i\theta}$.
We consider a rotating frame, so that the complex amplitude of the LO is constant.
By combining at a beamsplitter the LO and an instantaneous field represented by the annihilation operator $\hat{a}(t)$, we obtain the two output field amplitudes $\hat{a}_\text{out-A}(t)=(\alpha+\hat{a}(t))/\sqrt{2}$ and $\hat{a}_\text{out-B}(t)=(\alpha-\hat{a}(t))/\sqrt{2}$.
We then detect their photocurrents and take the difference of the two signals,
which is proportional to a certain quadrature,
\begin{align}
\hat{n}_\text{out-A}(t)-\hat{n}_\text{out-B}(t)
\approx & \alpha^\ast\hat{a}(t)+\alpha\hat{a}^\dagger(t) \notag\\
= & 2|\alpha|(\hat{x}(t)\cos\theta+\hat{p}(t)\sin\theta) \notag\\
\equiv & 2|\alpha|\hat{x}^\theta(t).
\end{align}
Therefore, by adjusting the phase $\theta$ of the LO, we can measure any quadrature at an arbitrary phase.

The quadrature values $\hat{x}_\psi$ of mode $\psi$ can be expressed as an integral over the instantaneous quadratures,
\begin{align}
\hat{x}_\psi\equiv & \frac{\hat{a}_\psi+\hat{a}_\psi^\dagger}{\sqrt{2}} \notag\\
=& \int dt\frac{\psi^\ast(t)\hat{a}(t)+\psi(t)\hat{a}^\dagger(t)}{\sqrt{2}} \notag\\
=& \int dt\{\Re[\psi(t)]\hat{x}(t)+\Im[\psi(t)]\hat{p}(t)\}.
\end{align}
Since we cannot simultaneously measure $\hat{x}(t)$ and $\hat{p}(t)$ at any time $t$, in general, we do not obtain a measurement outcome for $\hat{x}_\psi$ from the instantaneous outcomes.
However, if the mode function $\psi$ is a real function in the rotating frame of the LO, we obtain $\hat{x}_\psi$ by integrating over $\hat{x}(t)$ with a weight function of $\psi(t)$.
This condition is the same as matching the frequencies of the signal mode and the LO.

The quadrature distribution of an arbitrary single-mode state $\hat{\rho}^{\psi_0}=\sum_{m,n}c_{m,n}\ket{m}\bra{n}$ is sensitive to the phases $\theta$.
Such a state can be reconstructed from the distributions of the quadratures $\hat{x}_{\psi_0}^\theta$ for various phases $\theta$.
However, all photon-number states, including the vacuum and single-photon states as well as their incoherent mixtures, have phase-insensitive quadrature distributions, which exactly applies to our experiment.
Therefore, we could reconstruct a diagonal density operator $\hat{\rho}^{\psi_0}=\sum_{n}c_{n}\ket{n}\bra{n}$, containing the corresponding photon-number distribution, from the quadrature distribution for some single phase.
As described in Sec.~\ref{sec:experiment} of the main text, the phase-insensitivity of our quantum states was experimentally
confirmed by checking that two distributions coincide:
one for which the LO phase $\theta$ was slowly scanned, and another one for which it was not scanned.
This phase scan was slow enough so that the phase remained constant within each photon pulse.
In the following, we choose the measured observable to be $\hat{x}$ without loss of generality.

As can be found in standard textbooks, the quadrature distributions of the vacuum and single-photon states are,
\begin{align}
P(\hat{x}=x;\ket{0}\bra{0}) = & \frac{1}{\sqrt{\pi}}\exp(-x^2) \equiv P_0(x), \notag\\
P(\hat{x}=x;\ket{1}\bra{1}) = & \frac{2x^2}{\sqrt{\pi}}\exp(-x^2) \equiv P_1(x).
\label{eq:marginal_distribution}
\end{align}
Similarly, all multi-photon states have their own distributions.
The distribution for an incoherent mixture of photon-number states can be decomposed into distributions of pure photon-number states,
\begin{align}
&P(\hat{x}=x;\sum_nc_n\ket{n_\psi}\bra{n_\psi}) \notag\\
&\quad = \sum_nc_nP(\hat{x}=x,\ket{n_\psi}\bra{n_\psi}).
\end{align}
The photon-number distribution corresponds to the coefficients of this decomposition.
In the decomposition of the experimental distribution, we maximized the likelihood by using the downhill simplex method.

\subsection{Mode Function Estimation by Principal Component Analysis}

As discussed above, if we know a priori the real mode function $\psi$ of the single photons, we can obtain the density operator $\hat{\rho}^{\psi_0}$ that maximizes the purity $\bra{1}\hat{\rho}^{\psi_0}\ket{1}$ by setting $\psi_0 = \psi$.
In the experiment, we estimated the mode function $\psi$ from the homodyne raw data.
This was possible, because only a single mode $\psi$ was excited, and the other orthogonal modes were basically in vacuum states.
Therefore, we could easily observe the target mode $\psi_0$, since its photon number is maximized.
Note that this maximization of the photon number is equivalent to the maximization of the phase-insensitive quadrature variance,
\begin{align}
2\langle\hat{a}_{\psi_0}^\dagger\hat{a}_{\psi_0}\rangle
= \langle\hat{x}_{\psi_0}^2\rangle + \langle\hat{p}_{\psi_0}^2\rangle - 1.
\end{align}

From the homodyne raw data, we can calculate the auto-covariance,
\begin{align}
V_{t,t^\prime}\equiv\langle\hat{x}(t)\hat{x}(t^\prime)\rangle.
\end{align}
Then, optimization of the mode function $\psi_0$ corresponds to finding the eigenfunction of the auto-covariance, which then gives the maximal eigenvalue.
Therefore, the problem is solved via the spectral decomposition of the auto-covariance.
In the experiment, owing to the finite sampling rate and the finite frame length, the auto-covariance gives a finite-dimensional square matrix, whose diagonalized form serves our purposes.
This method of decomposing a mixed signal into independent signals is referred to as the principal component analysis (PCA).

Note that, in order to validate this method, the detector response must be sufficiently flat in the frequency region of interest.
In addition, the PCA never discriminates two components with the same variance.
Therefore, in case there is some noise due to experimental imperfections, with a variance close to that of a single-photon quadrature, the resulting function becomes in general a combination of that noise and the true function.

The quadrature samples are random, obeying some statistics.
Therefore, an estimation via the PCA may capture the accidental deviation from the true distribution, which increases the eigenvalue, leading to some overestimation of the average photon number.
This query then affects the precision of the estimation of the mode function $\psi_0$.
We checked that the mode match $|(\psi_0,\psi_0^\prime)|^2$ between two mode functions obtained from two different sets of quadrature samples was typically more than $99\%$.
Even when we used different sets of samples for the estimation of mode $\psi_0$ and for the estimation of the photon-number distribution $\hat{\rho}^{\psi_0}$, the change in the obtained purity was negligible.

\subsection{Wigner Function}

A Wigner function $W(x,p)$ of a single-mode quantum state is a two-dimensional quasi-probability distribution in phase space, characterized by an operator algebra $[\hat{x},\hat{p}] = i$,
\begin{align}
W(x,p) = \frac{1}{2\pi}\int_{-\infty}^{\infty}\exp(ip\eta)\langle x-\frac{\eta}{2}|\hat{\rho}|x+\frac{\eta}{2}\rangle d\eta.
\end{align}
It resembles a classical distribution function $W_c(x,p)$ in the phase space of a general position $x$ and momentum $p$, where each point of the function $W_c(x,p)$ has a meaning of a joint probability density.
Just like a classical distribution function, the Wigner function gives the correct quadrature distributions via its marginal distributions,
\begin{align}
\int_{-\infty}^\infty W(x,p)dp =& P(\hat{x}=x), \\
\int_{-\infty}^\infty W(x,p)dx =& P(\hat{p}=p).
\label{eq:WigMarginal}
\end{align}
The distribution of a quadrature $\hat{x}^\theta$ at an arbitrary phase $\theta$ is obtained similarly.

The biggest difference between Wigner and classical distribution functions is that Wigner functions allow for negative values.
Due to the uncertainty principle, we can never determine the conjugate variables $\hat{x}$ and $\hat{p}$ simultaneously.
Therefore, a single point of the Wigner function $W(x,p)$ alone cannot be interpreted as a joint probability; only when integrated, the meaning of a probability density arises, like in Eq.~\eqref{eq:WigMarginal}.
For this reason, Wigner functions may take on negative values -- a strong quantum feature, because it is a consequence of the noncommutative operator algebra.
Note that even though $W(x,p)$ may take on negative values, the integrated probability densities never become negative.

In fact, a single-photon state is a prominent example of such nonclassical states with a negative Wigner function.
The Wigner functions for a single-photon and a vacuum state are given by,
\begin{align}
W_{\ket{0}\bra{0}}(x,p)
=& \frac{1}{\pi}\exp(-x^2-p^2), \\
W_{\ket{1}\bra{1}}(x,p)
=& \frac{1}{\pi}(2x^2-2p^2-1)\exp(-x^2-p^2).
\end{align}
As expected, they are rotationally symmetric in the $x$-$p$ plane.
The function is always positive for a vacuum state (being a Gaussian function),
whereas it has a negative region for a single-photon state.
In particular, at the phase-space origin, the function value is $W_{\ket{0}\bra{0}}(0,0) = 1/\pi$ for a vacuum state, while it is $W_{\ket{1}\bra{1}}(0,0) = -1/\pi$ for a single-photon state.
Since the map between a Wigner function and a density operator is linear, the Wigner function for an incoherent mixture of a vacuum state and a single-photon state is obtained by summing their Wigner functions with the corresponding weights,
\begin{align}
&W_{p\ket{1}\bra{1}+(1-p)\ket{0}\bra{0}}(x,p) \notag\\
&\quad = pW_{\ket{1}\bra{1}}(x,p) + (1-p)W_{\ket{0}\bra{0}}(x,p).
\end{align}
Thus, if the single-photon component exceeds one half, the total Wigner function has negative values around the origin.
More precisely, for even-photon-number states, we have $W_{\ket{2k}\bra{2k}}(0,0) = 1/\pi$, while for odd-photon-number states, $W_{\ket{2k+1}\bra{2k+1}}(0,0) = -1/\pi$.
Therefore, if the contribution of multi-photon components, e.g., a three-photon component, is no longer negligible, the Wigner function can take on negative values even if the single-photon purity is below one half.

\end{document}